# Free energy landscape of sodium solvation into graphite

Ali Kachmar*[1], William A. Goddard III*[2]

[1]Qatar Environment and Energy Research Institute, Hamad Bin Khalifa University, Qatar Foundation, Doha, Qatar

[2]Materials and Process Simulation Center, California Institute of Technology, Pasadena, California 91125, United States

E-mail: akachmar@hbku.edu.qa, wag@wag.caltech.edu

## Abstract

Na is known to deliver very low energy capacity for sodium intercalation compared to Lithium. In this study, we use quantum mechanics based metadynamics simulations to obtain the free energy landscape for sodium ion intercalation from Dimethyl sulfoxide (DMSO) solvent into graphite. We find that the lowest free energy minima from the metadynamics are associated with sodium solvated by 3 or 4 DMSO. The free energy minima of these states are activated by a free energy of solvation computed to be 0.17 eV ($\Delta G(Na^+@(DMSO)_4$ - $\Delta G(Na^+@(DMSO)_3 \sim 6.6\ k_BT$), which in turn are the most thermodynamically stable. We observe weak interactions of sodium with graphite sheets during the unbiased and biased molecular dynamics simulations. Our simulations results suggest that solvent plays an important role in stabilizing the sodium intercalation into graphite through shielding of the sodium, and also from the interaction of the solvent with the graphite sheets. This suggests that the poor performance of Na is because the nonbonding and maybe partial covalent bonding of $Na^+$ to the DMSO is too strong compared to insertion into the graphite. This suggests that we consider solvents containing oxygen groups that might interact with Na more compatible with the bonding of Na in the GIC, but with negative charges (i.e., charge carrier nature) attach to these groups. In order to facilitate this intercalation, we propose solvents with negatively charged groups and aromatic cores (e.g., cyclic ethers) that could allow a greater rate of anion exchange to increase $Na^+$ mobility.

## I. Introduction

The Grand Challenge problem in batteries and electro-catalysis is elucidating the nature of the Electrode-Electrolyte or Solid-Electrolyte Interface (the EEI or SEI). Sodium Ion Batteries (SIBs) are the most cost-effective alternative to current generation of lithium ion batteries (LIBs) [1-3]. Historically, the first experiments on lamellar compounds of sodium with graphite were carried out by R. C. Asher, who suggested that lamellar graphite compounds could be intercalated with sodium although its stability is low [4-5]. In recent years, a tremendous amount of research has been devoted to sodium insertion, intercalation, and co-intercalation into disordered carbons and graphite for their application in solid-state electrochemical devices such as batteries [6-10]. The reactivity of sodium with carbonaceous materials depends on the nature of the carbon, and on the reaction conditions. Owing to its crystal and electron band structures, graphite constitutes one of the most important host lattices for intercalation of neutral and charged chemical species. Graphite is known to be a cathode for lithium intercalation with a theoretical capacity of a 372 mA g$^{-1}$. The electrochemical intercalation of Li$^+$ into graphite remains a subject of interest for capacitors and for energy storage. Soduim Graphite Intercalation Compounds (GICs) are formed with stage transformations [11,12], and very extensive studies have been devoted to investigate their physical and chemical properties [6]. The first stage of a two-dimensional sodium graphite intercalation compound was reported by Udod and Gé [11,12]. However, graphite is still not considered to be a good cathode for sodium ion battery due to the low energy density or capacity. Particularly for batteries the question is how the interaction of the Na or Li cation with the electrolyte transforms to a metallic Na, say, for the Na metal or Na Graphite Intercalation Compounds (GIC). Intercalation of other alkali metals such K$^+$ or alkaline-earth metal like Mg$^{2+}$ as the active cations into graphite for storage capacity has attracted interest into using electrochemical methods and first principles simulations [13-17] to understand their intercalation behavior. Recently, Komaba et al. succeeded in intercalating K$^+$ into graphite, finding intercalation results for forming stage-1 KC8 compound that deliver good reversible capacity retention the battery (224 mA g-1) [16]. Y. Cao et al. succeeded inserting sodium (Na$^+$) in hollow carbon nanowires used as anode material, which delivers a high reversible capacity of 251 mAh/g over 400 charge-discharge cycles [17]. Y. Liu et al. identified the origin of weak sodium capacity in graphite than lithium, which had been a longstanding puzzle [15]. Co-solvent intercalations into GICs is another interesting topic which targets the understanding of the electrochemical and chemical reaction of the co-intercalated species. It is known that co-intercalation of the solvent molecules with the lithium during the cathodic reduction of graphite leads to a strong swelling of the carbon layers and thus reduces the van der Waals inter-layer energy, which stabilizes the graphite lattice. This was demonstrated by the early work of Besenhard et al. on the alkali metals and the tetra alkyl-ammonium cations intercalation into graphite in DMSO [18,19], and in DME [19] upon the carbon cathodic reduction to form a so formed ternary GICs of general composition Cn$^-$Li$_x$(solv)$_y$. Similarly, it was reported in early 1980 that in solvent stable to reduction such as DMSO and DME, solvated sodium is readily intercalated into graphite and forms a stage-1 ternary GICs [11,13,18,20-22].

In the last few years, many interesting experimental and theoretical studies have examined Na$^+$, Li$^+$, and K$^+$ intercalation assisted by various kinds of solvent [23-33]. We want to establish here if co-solvent intercalation phenomena can provide an alternative way to stabilize Na$^+$ into GIC. Recently, Kim et al. observed the first sodium stage-1 using *in operando* x-ray diffraction analysis for solvated-Na-ion intercalation with glyme (i.e., solvent containing oxygen atoms) into graphite, and reported that the intercalation potential shifts to higher values with increasing the chain length [27-29]. The later was also observed by Jache et al [29]. The formation of various stagings with solvated-Na-ions in graphite was also observed, and precisely quantified [27,29]. Yet, experiments have not been able to probe the 1-2 nm region that is essential for the EEI or SEI, and previous molecular dynamics simulation have nearly all used force fields, which forces a particular nonreactive model on the description. Clearly the way to make progress is to use first-principles simulations but a proper description of this interface requires 1000's of atoms for 1 ns or more, which is not practical. Here the most difficult to understand is how the metal ion can change its coordination with the solvent in order to be positioned for insertion into graphite or at metal surface. We take up this challenge here. Na is known to deliver very low energy capacity for sodium intercalation compared to Lithium. In order to understand the unusual behavior of sodium intercalation into graphite (i.e., the low sodium capacity battery), we investigate here atomistic level simulations of the mechanism for solvent assisted sodium intercalation into graphite, as a first step in understanding how the local environment can affect the free energy of solvation.

Our aim in this work is to determine the free energy of solvent co-intercalation with sodium into GICs. To do this, we used first-principles simulations aided by a metadynamics approach (MTD) to enhance the sampling of all possible sodium solvation states to obtain the free energy for sodium solvation. We first discuss the structure of the sodium local environment co-intercalated into graphite including radial distribution functions, sodium solvation structure, and coordination. We then analyze and discuss the structure and the thermodynamic stability of the various solvation scenarios for co-intercalated sodium with Dimethyl Sulfoxide (DMSO) and trifluoromethanesulfonimide (TFSI) oxygen-containing solvents into GIC. We show that the most favorable sodium solvation scenarios are associated with coordination of three or four DMSO solvent molecules. We discuss also other local solvation environments by predicting their free energy barriers, and their thermodynamic stability along with the minimum energy pathways connecting the most probable local minima.

## II. Computational details
### i. Static and molecular dynamics simulations

Simulations were carried out using the CP2K code [34]. We also used the hybrid Gaussian and Planewave method (GPW) as implemented in the Quickstep module of the CP2K package [35,36]. The control of the temperature was implemented for the ionic degrees of freedom by using Nosé-Hoover thermostats with 3 chains. This used a target temperature of 300 K and a time constant of 50 fs [37,39]. The electronic structure properties were calculated using the PBE functional [40] including the Grimme empirical correction scheme (DFT-D3) to include into the static and molecular dynamics (MD) simulations the dispersion interactions (vdW attraction) essential in graphite and for anion and cation solvation into graphite sheets [41,42]. The Kohn-Sham orbitals are expanded in a Gaussian basis set (MOLOPT-DZVP-SR-GTH) for all atom calculations for Na, C, S, O, and H. We employed the norm-conserving GTH pseudopotentials [43]. The auxiliary planewave (PW) basis set was defined by the energy cutoff of 550 Ry, and a relative cutoff was set to 70 Ry. For the case of static and dynamics calculations over the $Na^+[TFSI]^-[DMSO]_4$ cluster that we placed it into graphite, we used the Martyna-Tuckerman Poisson solver (MT) [44] to achieve a good description of the wavefunction in the non-periodic direction. Recent experimental study combined with first principles simulations shows a stability improvement of $Na^+$ in NaTFSI/DMSO solution, which motivates us to investigate the stability of $Na^+$ co-intercalation into graphite assisted by the same solvent. And this study revealed that the sodium has on average a fivefold-coordination [45], suggesting the short-range interactions between the sodium and the electrolytes are dominant. Another reason we take this particular solvent mixture is motivated by previous solvation free energy calculations of sodium into DMSO ranging from 0 to 5 molecules [46] to use the results as a baseline for our free energy of sodium solvation into graphite, but also relied on previous experimental data for the lithium co-intercalation into graphite with similar solvent species [23].

The model GICs system labeled as $C_n^-Na_x(solv)_y$ (i.e., $Na^+(sol)@GIC$) is based on $C_{16}$ which was simulated by a 4x4x1 hexagonal super-cell including 2 graphite sheets consisting of 128 C atoms each, and $Na^+$ co-intercalated with a $[TFSI]^-[(DMSO)_4]$ cluster into the graphite sheets. This contains one Na atom, one TFSI as counter ion to keep the system neutral, and 4 DMSO solvent molecules. The system contains 256 carbon atoms for the graphite sheets, and 56 atoms for the solvated sodium with the DMSO and TFSI solvent molecules. The cell size of the $C_n^-Na_x(solv)_y$ model was optimized to allow for the rotation and translation of the solvent molecules surrounding the sodium cation. We found an interlayer spacing ($d_s$) of 13.30 Å to accommodate the solvated sodium with the 4 DMSO and one TFSI molecules, including the rotation/translational motion of the DMSO and the TFSI during unbiased/biased molecular dynamics simulations. This model was used in order to have a good compromise between system size and computing resources, but also relied on previous experimental data for the lithium co-intercalation into graphite with similar solvent species [23]. To comment on the $d_s$, recent experiments and modeling by Kim al [27], proposed three staging models of one, two, and triple complex intercalation of $[Na-DEGDME]^+$ with graphene $d_s$ of 7.43 Å, 11.98 Å, and 16.40 Å respectively. This suggests that the $d_s$ is correlated to the staging models of GICs, and the concentration of the solvent molecules increases the $d_s$ to ease the de-solvation of the sodium cation. Goktas et al. modeled $Na^+(diglyme)_2$ complex intercalated into graphite with a d-spacing of 11.3 Å. It appears from all the previous studies that the interlayer spacing really depends on the size, the shape and the number of the co-intercalated solvents, which are still a matter of debate in the community [47,48]. Periodic Boundary Conditions (PBC) was applied using a cell dimension of 17.140 x 19.792 x 17.264 Å$^3$. The time step of the integration of the dynamic equations was set to 1 fs. The simulations were extended up to 100 ps. Trajectories were saved every 1 fs. Singlet spin states were used in all calculations. The Brillouin zone was sampled at the Gamma point only throughout the present study. Finally, we have used the VMD and Molden software for structures visualization [49,50].

### ii. Metadynamics simulations

The production trajectory was generated by running 100 ps in the NVT ensemble at 300 K, followed by more than 356 ps using the MTD acceleration scheme on the model system $C_n^-Na_x(solv)_y$ that contains 312 atoms [51,52]. MTD was used to sample over the sodium solvation phase space into GIC in order to investigate the various sodium solvation scenarios with the solvent molecules and to obtain their free energies. The initial 100 ps NVT of the model system ($Na^+[TFSI]^-(DMSO)_4$) co-intercalated into GICs was used to define and calibrate two collective variables (CVs) that describe the local structure, to observe the changes in coordination, and the de-solvation from the solvent molecules surrounding the sodium cation. The first, $CV_1$, was the coordination between the sodium cation ($Na^+$) and the oxygen atoms of the DMSO solvent molecules, which provides sufficient information about the number and relative orientation of the DMSO molecules solvating the $Na^+$ cation. The second, $CV_2$, was the coordination between the sodium cation and the oxygen atoms belonging to the TFSI anion ($CV_2$), which provides also information about the sodium solvation and the rotation of the TFSI. The height and the width of the Gaussian are $3 \times 10^{-3}$ and 0.05 in a.u, respectively. The maximum number of MD steps between spawning two hills was set to 50.

## III. Results and discussion
### i. Structure and dynamics properties
#### a. Unbiased MD simulations: $C_n^-Na^+(sol)$ (i.e., $C_n^-Na^+(TFSI)^-(DMSO)_4$ )

We first optimized the cell and the positions of the $C_n^-Na^+(sol)$ model system to obtain a good guess for the molecular dynamics sampling of the $Na^+$ solvation between the two graphite sheets. The production trajectory was generated by running 104 ps in the NVT ensemble at 300 K. We discuss hereafter the structure of the sodium local environment into the $C_n^-(sol)$ model system described previously in section II.i. Through the MD simulations, we observe that the dominant interactions are between the sodium and the oxygen ligands in the solvent molecules. To better characterize the local environment of the sodium cation, we computed the g_DMSO, and g_TFSI pair distribution functions (g_DMSO(r): ($Na^+$-O (DMSO)) and g_TFSI(r): ($Na^+$-O (TFSI)) are: atom-atom pair distribution functions between the $Na^+$ and the O ligands in the DMSO and TFSI solvent molecules), and the solvation number of DMSO and TFSI molecules toward the $Na^+$ was evaluated from the running coordination numbers (n(r)). In Figure 1, we plot and compare the g(r) and n(r) between the $Na^+$ and the oxygen ligands in the solvent molecules. The g(r)'s for both g_DMSO(r) and g_TFSI(r) show a first peak at 2.36 Å, and a second one at 2.46 Å respectively. These peaks are characteristic of sodium interaction with the oxygen ligands in the DMSO and the TFSI solvent molecules. However, the cumulative radial distribution function between the $Na^+$ and the oxygen ligands in the DMSO solvent molecules has an average of 3 DMSO in the first solvation shell, and one additional DMSO molecule in the second solvation shell of sodium (see Fig.1b). Regarding the interaction between the $Na^+$ and the oxygen ligands in the TFSI molecule, the g_TFSI(r) shows a first peak at 2.46 Å slightly shifted from the $Na^+$- O (DMSO) distance, and a second peak at a distance of 4.57 Å. Integrating to the first minimum we find that the sodium is coordinated to one oxygen ligand in TFSI.

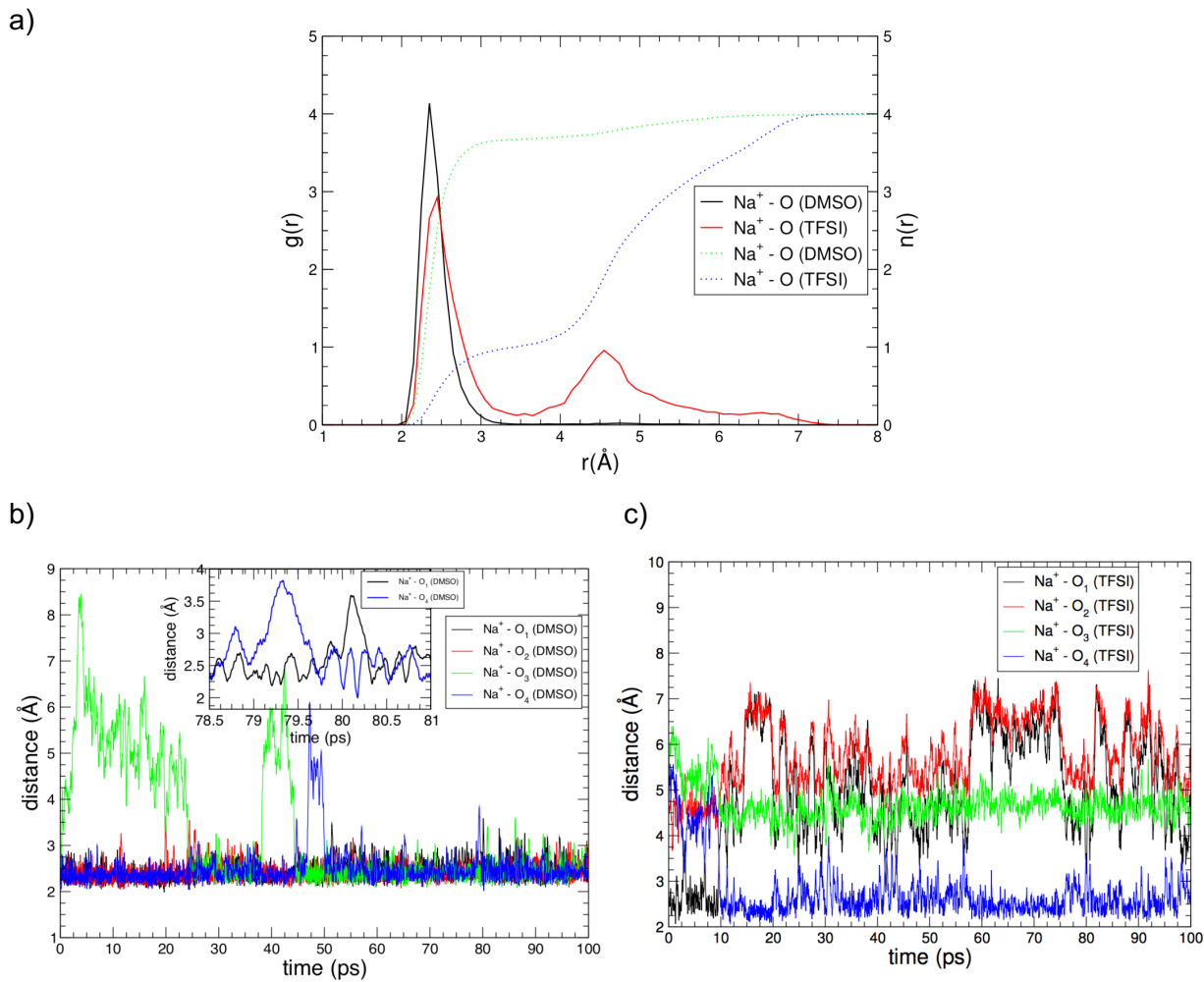

Figure. 1 a) The Na$^+$-O solvent radial distribution functions [solid lines] g_DMSO(r) and g_TFSI(r) with their corresponding cumulative functions n(r) [dashed lines]. b) and c) The distances between the Na$^+$ and the oxygen ligands in the DMSO and TFSI solvent molecules as function of the simulation time in picosec. b) The inset shows the distance between the Na$^+$ and O in the DMSO in the time domain 78 ps ≤ t ≤ 81 ps.

The analysis of the evolution of the distances between the sodium and the oxygen ligands in the DMSO molecules is interesting because it shows how the DMSO solvent molecules exchange within the sodium solvation shell. In Fig. 1b the first solvent exchange occurred between the time period 40 and 52 ps (the most pronounced transition from 4 to 3 coordination), where the solvent molecule 3 (labeled as $O_3$ (DMSO)) that was outside the solvation shell exchanged with the solvent molecule 4 (labeled as $O_4$ (DMSO)). The same dynamical exchange of solvent happened between $O_1$ (DMSO) and $O_4$ (DMSO) between the time domain of 78 and 81 ps. Zooming in in the time domain larger than 70 ps we could see other solvent exchanges between other solvent molecules. However, the *cis to trans* rotation of the TFSI in the first 10 ps in Fig. 1c from Na$^+$ - $O_1$ (TFSI) to Na$^+$ - $O_4$ (TFSI)) contributes to the Na$^+$ coordination reflecting higher flexibility and degree of freedom compared to the DMSO. This should result in promoting the sodium diffusivities into graphite through the electrostatic interactions between the sodium and the anion. As a side note, we show only the g(r)'s and n(r)'s between the Na$^+$ and the oxygen ligands in the solvent molecules because we are mainly interested in the sodium solvation processes.

a) 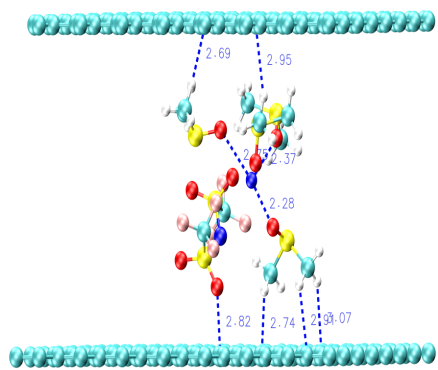

b) 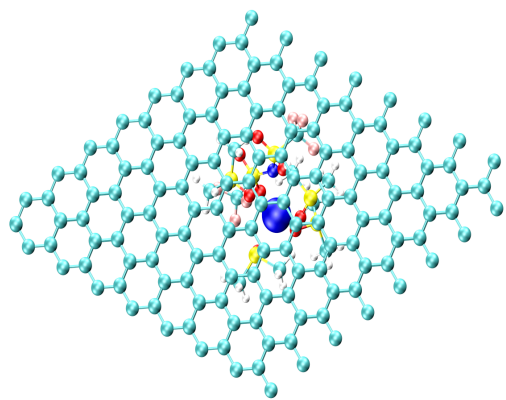

c) Na$^+$ - (DMSO)$_3$

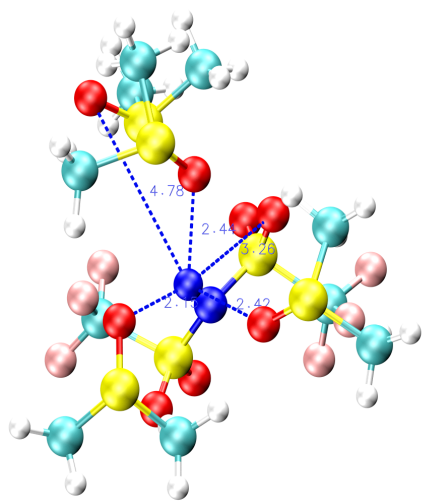

d) Na$^+$ - (DMSO)$_4$

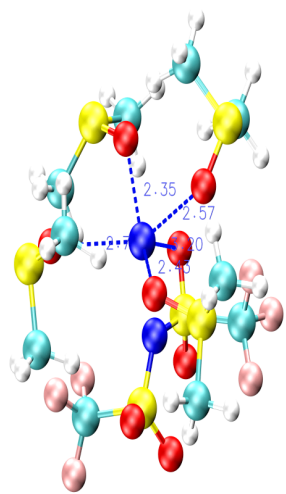

e) Na$^+$ - (DMSO)$_3$(TFSI)

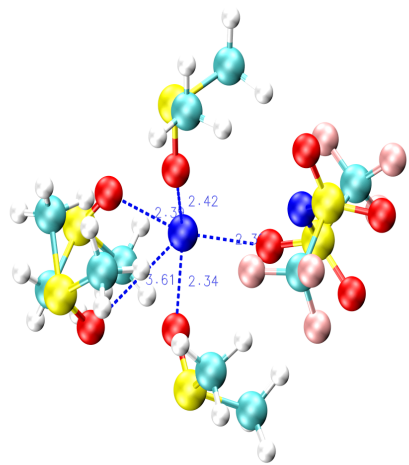

Figure 2 Snapshots of sodium co-intercalated with solvent in between the graphene sheets a) side view, and b) top view respectively. Snapshots c), d), and e) are selected from the unbiased molecular dynamics trajectory representative of the sodium coordination with 3, 4 DMSO and a mixture between DMSO and TFSI solvent molecules. These snapshots are given within a 3 Å radius of the sodium cation. Distances between the Na$^+$ and the oxygen ligands in the solvent molecules are shown with blue dashed lines. The two graphite sheets are omitted for the sake of clarity. Atom colors: Na$^+$ (blue), O (red), F (pink), S (yellow), C (green), H (white) and N (blue).

### b. Biased MD (MTD) simulations results: $C_n^-Na^+(sol)$

We follow the approach used recently to map the free energy of lithium solvation in a protic ionic liquid [53], which provided meaningful information about the possible lithium solvation scenarios in the protic ionic liquid Ethyl-Ammonuim Nitrate. We use the same approach in this study to map the free energy landscape of sodium co-intercalation with solvent into GICs (i.e., $C_n^-(sol)$). To better characterize the dynamics of the sodium solvation during the biased molecular dynamics, we report in Figure 3 the evolution of the collective variables ($CV_1$, and $CV_2$) as function of time along the MTD simulations. We note here that the changes in sodium coordination ($CV_1$ or $CV_2$) behaves similarly to the observations in the unbiased molecular dynamics simulations. This was characterized by the structure analysis reported in section III.i(a). The $CV_2$ fluctuates faster than the $CV_1$ because of the higher degree of freedom of the TFSI compared to the DMSO. It also highlights that the Na$^+$ interaction with TFSI is weaker than the interaction with DMSO. The fluctuations of $CV_1$ are in the range ($0.5 \leq CV_1 \leq 4$) while $CV_2$ is between ($0 \leq CV_2 \leq 3$), as can be seen in Figure 3.

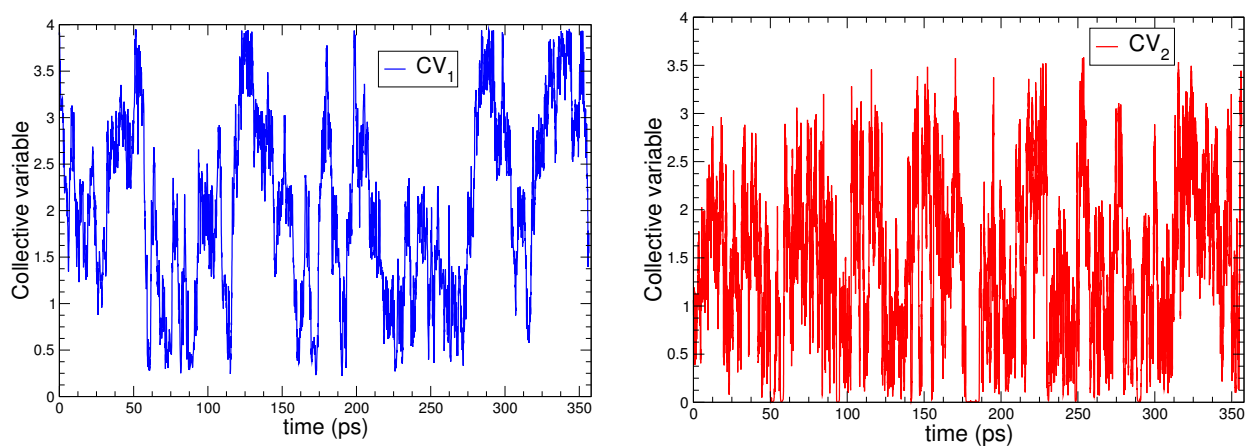

Figure. 3 Collective variables $CV_1$ (left hand side) and $CV_2$ (right hand side) along the simulations trajectory.

The collective variables never find complete sodium de-solvation in either the MTD simulations or in the non-biased MD simulations. However partial sodium de-solvation does occur. Moreover, we see a displacement of the anion (TFSI) and solvent molecules exchange within the sodium solvation shell (see Fig. 1b, Fig. 1c)). This was also observed by Yamada et al. [23]. These results suggest that the interaction of Na$^+$ with DMSO is much stronger than that of TFSI. The de-solvation energy of sodium (that correspond to the low sodium coordination, $CV_1$=0.5 or 1 DMSO molecule) seems to be very high, which suggests that the strength of sodium solvation is more important than lithium solvation in graphite. In conclusion, the biased molecular dynamics shows a dynamical behavior of the sodium (Na$^+$) local environment that is nicely reproduced in the unbiased MD simulations reported in section III.i(a).

### ii. Free energy

MTD with the defined collective variables ($CV_1$, $CV_2$) in the section II.ii was carried out to characterize all possible solvation scenarios of sodium co-intercalation into $C_n^-$(Sol) w.r.t to their free energy barriers. We show in Figure 4 a two-dimensional (2D) free energy landscape constructed from the MTD as function of the collective variables $CV_1$ (x-axis), and $CV_2$ (y-axis). From the 2D map, we see that the sodium explores different solvation states. The lowest energy minima in the phase space, which are more frequently visited by the MTD are associated with sodium solvation with only the oxygen ligands of DMSO solvent molecules. The first lowest free energy of solvation corresponds to a *threefold* $Na^+$ coordination with DMSO (i.e., $CV_1=3$), with the free energy for the sodium to be in this state computed to be -1.86 eV. The second lowest energy minima associated with a sodium coordinated with DMSO molecules but with a lower coordination number (i.e., n(r)= 2.5) along the $CV_1$, leading to a free energy of 0.035 eV higher than the lowest energy minima (i.e., $CV_1=3$). The thermodynamic stability for a change in sodium coordination between two different coordination's is defined as the difference between their free energy as $\Delta G^{difference}$ ~ [$\Delta G[Na^+(DMSO)_x] - \Delta G[Na^+(DMSO)_{x+1}]$, where x is the number of the DMSO molecules in the sodium solvation shell]. Thus, these two minima ($CV_1=2.5$ and $CV_1=3$) are thermodynamically stable and thermally activated at room temperature. The sodium in these two configurations is observed to transfer from one to the other, suggesting that these configurations belong to the same basin. However, along this collective variable there are also some interesting solvation states to be discussed. These states are associated with the sodium coordination with DMSO but with a coordination number (CN) of 2 and 4 DMSO molecules. The 4 coordination leads to a higher energy with respect to the lowest one ($CV_1=3$) with a free energy difference of ~ 0.17 eV ~ 6.6 $k_BT$. Therefore, threefold and fourfold $Na^+$ coordination with DMSO are thermodynamically stable and the $Na^+$ can transfer from one to the other and vis versa with an energy barrier of -0.17 eV ~ 6.6 $k_BT$.

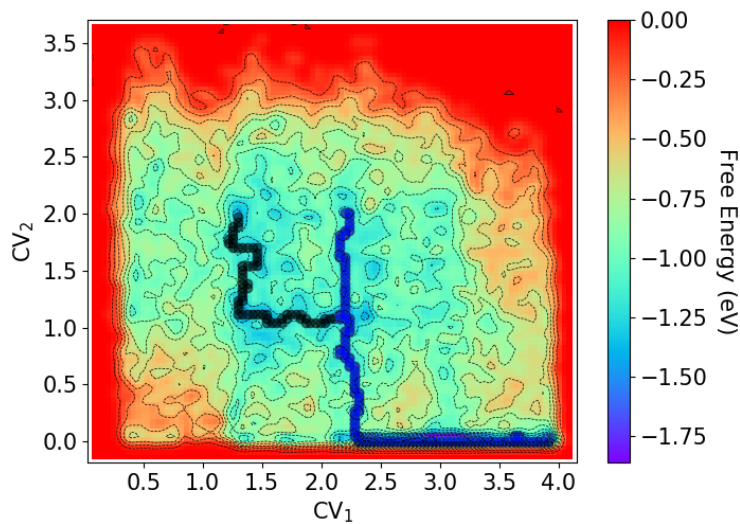

Figure. 4 Free energy surface as estimated from the MTD potential acting in the reduced phase space of the two selected collective variables. The contour lines are reported with a spacing in energy of 0.01 eV, the color map distinguish more probable areas (blue) from less visited areas (orange-red). The black and the blue dots mark the two identified Minimum Energy Pathways for the sodium (MEP's) [54].

In the regions of low coordination in free-energy space, the sodium shows that coordination's of 1 and 2 are also accessed but they are not thermodynamically favorable. An example, the lowest sodium ($Na^+$) coordination ($CV_1=1$; $CV_2=0$) is a state that is higher in energy and thermodynamically not stable with a free energy barrier of -0.55 eV. This is one of the configuration where the sodium is moving towards or getting closer to one of the graphite sheets. Other possible low coordination sodium solvation states in free-energy space contain a mixture of oxygen ligands from the DMSO and TFSI solvent molecules but they are not thermodynamically favorable. As an example, a sodium coordination of two solvent molecules with only one oxygen ligand from DMSO, and a second oxygen ligand from the TFSI. The free energy in the low coordination regions ($CV_1 < 1$) are in the range of ~ -0.55 and -0.63 eV, and about -1.86 eV in the region of the well-defined minima that corresponds to the $CV_1=3$. As an example, the free energy is -1.46 eV for ($CV_1=1.5$, $CV_2=1.0$), and –1.42 eV for ($CV_1=1.5$, $CV_2=2.0$). Therefore, we list in Table 1 only the lowest Free Energy Minima (FEM) of the sodium solvation states that are in the range of: -1.4 eV ≤ FEM's ≤ -1.8 eV. The sodium CN's for both oxygen ligands (CN: O (DMSO)) and (CN: O (TFSI)) for all the lowest FEM extracted from the 2D free energy map are summarized in the Table 1.

Table. 1 $Na^+$ solvation states with their corresponding free energies

| Minimum | CN: ($Na^+$..O (DMSO)) | CN: ($Na^+$..O (TFSI)) | Free energy [eV] |
|---|---|---|---|
| $FEM^1$ | 3.0 | 0.0 | -1.862 |
| $FEM^2$ | 2.5 | 0.0 | -1.826 |
| $FEM^3$ | 4.0 | 0.0 | -1.692 |
| $FEM^4$ | 1.5 | 1.0 | -1.457 |
| $FEM^5$ | 3.5 | 0.0 | -1.455 |
| $FEM^6$ | 1.5 | 2.0 | -1.415 |
| $FEM^7$ | 2.0 | 1.0 | -1.411 |
| $FEM^8$ | 2.0 | 0.0 | -1.357 |

We identified two Minimum Energy Pathways (MEP's) in the 2D free energy map (Figure 4). The first one starts from the coordination (CN: O (DMSO)=1.4, CN: O (TFSI)=2.0) and decreases in energy until the coordination (CN: O (DMSO)=1.4, CN: O (TFSI)=1.0). It then continues to explore regions of free energy space along the $CV_1$, which are mainly located in the lowest free energy regions (i.e., CN: O (DMSO)= 3, and 4). The second MEP starts as well from the same $CV_2$ value or CN: (CN: O (DMSO)=2.4, CN: O (TFSI)=2.0) and goes down to the sodium coordination with oxygen ligands of the DMSO and TFSI with a value of $CV_2$ and $CV_1$ (CN: O (DMSO)=2.4, CN: O (TFSI)=0.0) and then continues to explore the lowest free energy space similarly to the first pathway. The free energy for those coordination's [CN: O (DMSO)=2, 3, 4 and CN: O (TFSI)=0.0] follow this free energy trend: $\Delta G[Na^+(DMSO)_3] > \Delta G [Na^+(DMSO)_4] > \Delta G [Na^+(DMSO)_2]$.

From the free energy data reported in Table 1, the sodium coordination with 3, and 4 DMSO solvent molecules are lower in free energy, which is in line with the observations during the unbiased molecular dynamic simulations (section III.i(a)) that are characterized by the radial and the cumulative distribution functions in Figure 1. However, the coordination 2 seems to be a transition state connecting the Na$^+$ coordination to 3 and 4 DMSO solvent molecules, leading to an energy barrier of 0.3 eV for the sodium to transfer from the coordination 4 to the coordination 3. Moreover, the transfer of the sodium from the coordination 3 to 4 is less favorable with an energy barrier of 0.5 eV. Thus, the energy penalty to push away a DMSO from the sodium solvation shell is lower by 0.2 eV than value for the bringing it back to its solvation shell. In order to correlate further our free energy data for the sodium solvation with the data of the lithium co-intercalation along with DMSO into graphite previously reported [23], we first make few observations from the analysis of our data:
1) Displacement of the DMSO molecules by the TFSI contribute to the sodium coordination via its oxygen ligands similarly to observations for the lithium case [23]
2) Lower sodium solvation states are accessible but higher in free energy
3) The unbiased and biased MD simulations for the dynamical behavior of the sodium co-intercalation are similar

Experimentally, it has been reported that lithium solvation with three DMSO is a criterion for determining whether intercalation of lithium ion into graphite is favorable [23]. Yet, our biased simulations predict that sodium co-intercalation with 3 DMSO molecules (i.e., with $CN^{DMSO}=3$ or $CV_1=3$) is linked to the lowest free energy minima mapped by the MTD. Recently, M. Hue et al. combined Raman spectra and Ab initio molecular dynamics calculations to show that Na$^+$ solvation in 4.1 m TFSI/DMSO electrolytes forms a Na(DMSO)$_3$TFSI-like solvation structure [45]. This solvation structure has fivefold Na$^+$ coordination and is one of the configuration scanned by the MTD (i.e., $CV_1=3$, $CV_2=2$) but we find that it has a higher free energy minimum of -0.73 eV. We can also compare our calculations for the co-intercalation energy for sodium into graphite with DMSO, with some other data available in the literature for both Li$^+$ and Na$^+$ co-intercalation with diglyme [33]. Table 2 summarizes our data and the data available in the literature.

| Table. 2 Free energy of co-intercalation and intercalation energy for Li$^+$ and Na$^+$ into graphite (eV) | | | |
|---|---|---|---|
| Free energy of co-intercalation (Na$^+$-(DMSO)); this study | Free energy of co-intercalation (Na$^+$-(DMSO)$_3$: threefold); this study | Intercalation energy ([Li$^+$-diglyme]C$_{16}$: threefold) [33] | Intercalation energy ([Na$^+$-diglyme]C$_{16}$: threefold) [33] |
| -0.55 | -1.862 | -0.32 | -0.26 |

First, the free energy of co-intercalation for Na$^+$ coordination with one oxygen ligand in the DMSO molecule is computed to be -0.55 eV. The equivalent de-solvation energy (0.55 eV) falls in the range of sodium de-solvation energies (i.e., 0.42-0.73 eV), which are commonly small, mainly due to the weaker Lewis acidity of Na$^+$ ion than Li$^+$ ion [55]. It was reported that the solvation free energy of Lithium is lower than sodium for the coordination that goes from 0 to 5 DMSO solvent molecules [46]. Therefore, the free energy of co-intercalation into graphite for sodium as well for the lithium could follow the same trend. A fair comparison would be a threefold Na$^+$ coordination with two different solvents that fulfil a threefold coordination with Na$^+$ such as DMSO, and diglyme that was successfully co-intercalated with sodium into graphite [23,27,46].

We turn our attention now to the threefold sodium co-intercalation with diglyme and DMSO solvents. The free energy of co-intercalation of sodium (threefold coordination) with DMSO in our study is computed to be -1.862 eV while the intercalation energy of Na$^+$ with diglyme is reported to be -0.32 eV [33]. We believe that the main reason for this energy difference is the entropic contribution to the free energy. Thus, it is hard to say in which solvent the co-intercalation energy is more favorable since our free energy data ($\Delta G_s = \Delta H_s - T^*\Delta S_s$) contains the solvation enthalpy ($\Delta H_s$), and the entropic contribution ($-T^*\Delta S_s$) for a given sodium solvation state while the reported data does not. Nevertheless, our data confirm that the sodium ($Na^+$) is readily intercalated into graphite with DMSO which is in line with the earlier report by Yazami et al. [6,13], and shows that sodium has a stronger interaction with DMSO than diglyme. While the de-solvation energy of sodium is smaller than for lithium [46,55] in solvents such as DMSO, the sodium de-solvation state in graphite is believed to follow the same trend because it promotes the instability of the intercalation phenomena as predicted by the MTD simulations. This explains the low energy density of the sodium electrodes because a low activation energy implies faster charge/discharge, and therefore lower performance. However, a strong $Na^+$ solvation strength with DMSO implies a high free energy of de-solvation, and highly thermodynamically stable free energy minima. Yet, our study of co-intercalated sodium shows that solvation in graphite is stable with a lowest free energy minimum for a threefold sodium solvation with DMSO solvent molecules. This suggests, that the strength of $Na^+$ solvation is important in this particular solvent, and that this leads to slow kinetics compared to lithium, i.e., slow ionic transport properties. This indicates that the sodium diffusion constant in graphite is lower by two or three orders of magnitude ($D(Na^+) \sim 10^{-11}$ $cm^2s^{-1}$ [56,57]) than the measured and computed values for lithium ($D(Li^+) \sim 10^{-6}$ $cm^2s^{-1}$ [58]; $D(Li^+) \sim 6 \pm 1 \times 10^{-8}$ $cm^2s^{-1}$ [59]) and potassium ($D(K^+) \sim (6 \pm 2) \times 10^{-7}$ $cm^2s^{-1}$ [57]; 10 times bigger than the one for $Li^+$) respectively. The reason why it is so important for the alkali-metals to be free from any solvent when being intercalated into graphite is to achieve a reversible electrode behavior and high capacity. But this is not the case for sodium, which shows better stability upon solvation and instability upon de-solvation. It is known for co-intercalation to occur into graphite, the lithium or sodium needs a higher free energy of complexation, but for the cation to be de-solvated the free energy needs also to be high. A higher de-solvation or higher activation free energy is naturally/thermodynamically associated to a strong solvation strength. Overall, the free energy of low sodium coordination with the solvent appears to be accessible but thermodynamically unstable. Therefore, the solvent (i.e., solvation strength) plays an important role in stabilizing the sodium intercalation into graphite (i.e., the solvent screens the sodium from the graphite sheets). Thus, this finding could be the origin of the low capacity of graphite as a negative electrode for the SIB's.

## IV. Conclusion

We use first-principles based MD aided by the metadynamics simulations to analyze of the free energy landscape for sodium ion co-intercalation into graphite as assisted by solvent. We examined the relative stability between the different sodium solvation states that are thermodynamically favorable and stable. We found that the lowest solvation free energy of sodium is associated with threefold coordination with DMSO solvent. Interestingly, it was reported recently that lithium solvation with three DMSO is a criterion for its co-intercalation into graphite [23]. We conclude that solvent plays a major role in enhancing the intercalation of sodium into graphite. Our findings and previous findings on sodium co-intercalation with solvent, give clear evidence for a relationship between the strength of solvation versus the stability for alkali-metals and their intercalation behavior into graphite. Our study also supports recent experimental and theoretical work, opens a new way for sodium intercalation into graphite, and suggests the need to further investigate the sodium co-intercalation with screening other solvents to tune the electrochemical performance of graphite as a negative electrode for sodium ion battery. We propose the screening of solvent with lower tendency for sodium coordination, and with negatively charged groups like TFSI or solvent with aromatic cores like cyclic ethers, leading to delocalization of the negative charges. This should result in lower sodium coordination in the graphite and in enhancing the sodium intercalation and its diffusivities.

## V. Acknowledgements

A.K thanks his institution for the support to visit the Materials and Process Simulation Center, California Institute of Technology, Pasadena, CA. Warm thanks are due to Marcella Iannuzzi for stimulating discussions. The HPC resources and services used in this work were provided by the Research Computing group in Texas A&M University at Qatar. Research computing is funded by the Qatar Foundation for Education, Science, and Community Development. W.A.G. thanks Bosch Energy Research Network (BERN) for a Grant to support battery research.

## VI. Supporting Information (SI)

There are no SI to declare.

**TOC Graphic**

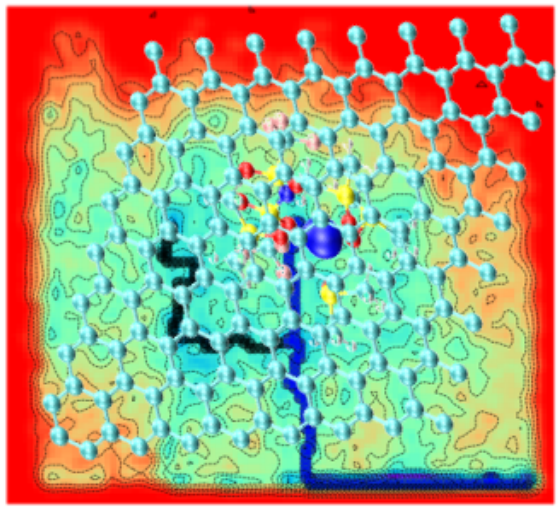

**Free energy landscape of sodium solvation into graphite**